\documentclass[floatfix,aps,prl,twocolumn,superscriptaddress]{revtex4}
\usepackage{graphicx}
\usepackage[]{epsfig}
\usepackage{times,amsmath,amssymb}

\usepackage{color}

\begin{document}

\title{Gate voltage dependence of noise distribution in radio-frequency reflectometry in gallium arsenide quantum dots}

\author{Motoya Shinozaki}
\affiliation{Research Institute of Electrical Communication, Tohoku University, 2-1-1 Katahira, Aoba-ku, Sendai 980-8577, Japan}

\author{Yui Muto}
\affiliation{Research Institute of Electrical Communication, Tohoku University, 2-1-1 Katahira, Aoba-ku, Sendai 980-8577, Japan}

\author{Takahito Kitada}
\affiliation{Research Institute of Electrical Communication, Tohoku University, 2-1-1 Katahira, Aoba-ku, Sendai 980-8577, Japan}

\author{Takashi Nakajima}
\affiliation{Center for Emergent Matter Science, RIKEN, 2-1 Hirosawa, Wako, Saitama 351-0198, Japan}

\author{Matthieu R. Delbecq}
\affiliation{Center for Emergent Matter Science, RIKEN, 2-1 Hirosawa, Wako, Saitama 351-0198, Japan}

\author{Jun Yoneda}
\affiliation{Center for Emergent Matter Science, RIKEN, 2-1 Hirosawa, Wako, Saitama 351-0198, Japan}

\author{Kenta Takeda}
\affiliation{Center for Emergent Matter Science, RIKEN, 2-1 Hirosawa, Wako, Saitama 351-0198, Japan}

\author{Akito Noiri}
\affiliation{Center for Emergent Matter Science, RIKEN, 2-1 Hirosawa, Wako, Saitama 351-0198, Japan}

\author{Takumi Ito}
\affiliation{Center for Emergent Matter Science, RIKEN, 2-1 Hirosawa, Wako, Saitama 351-0198, Japan}

\author{Arne Ludwig}
\affiliation{Lehrstuhl für Angewandte Festkörperphysik, Ruhr-Universität Bochum, D-44780 Bochum, Germany}

\author{Andreas D. Wieck}
\affiliation{Lehrstuhl für Angewandte Festkörperphysik, Ruhr-Universität Bochum, D-44780 Bochum, Germany}

\author{Seigo Tarucha}
\affiliation{Center for Emergent Matter Science, RIKEN, 2-1 Hirosawa, Wako, Saitama 351-0198, Japan}

\author{Tomohiro Otsuka}
\email[]{tomohiro.otsuka@riec.tohoku.ac.jp}
\affiliation{Research Institute of Electrical Communication, Tohoku University, 2-1-1 Katahira, Aoba-ku, Sendai 980-8577, Japan}
\affiliation{Center for Emergent Matter Science, RIKEN, 2-1 Hirosawa, Wako, Saitama 351-0198, Japan}
\affiliation{Center for Spintronics Research Network, Tohoku University, 2-1-1 Katahira, Aoba-ku, Sendai 980-8577, Japan}
\affiliation{Center for Science and Innovation in Spintronics, Tohoku University, 2-1-1 Katahira, Aoba-ku, Sendai 980-8577, Japan}

\date{\today}

\begin{abstract}
We investigate gate voltage dependence of electrical readout noise in high-speed rf reflectometry using gallium arsenide quantum dots. The fast Fourier transform spectrum from the real time measurement reflects build-in device noise and circuit noise including the resonator and the amplifier. We separate their noise spectral components by model analysis. Detail of gate voltage dependence of the flicker noise is investigated and compared to the charge sensor sensitivity. We point out that the dominant component of the readout noise changes by the measurement integration time.
\end{abstract}

\maketitle

Semiconductor quantum dots are artificial quantum systems we can control,\cite{1996TaruchaPRL,1997KouwenhovenScience} and  attracting much attention as possible building blocks of quantum computers\cite{1998DanielPRA} and research targets in condensed matter physics.\cite{1998GoldhaberNature, 2007HansonRMP} In the measurements of quantum dots, detection of single-electron charges in quantum dots with high-speed charge sensors is one of the important subjects for investigating quantum dynamics.\cite{2006FujisawaScience, 2006KoppensNature, 2014YonedaPRL} It is well known that quantum point contacts or quantum dots are useful as charge sensors to detect single-electrons in quantum dots.\cite{2014YonedaPRL, 2017OtsukaSciRep, 2019OtsukaPRB, 1991LafargeZPB, 1993FieldPRL} Because electrostatic coupling between the sensors and the target quantum dots acts as an effective gate voltage, the charge can be detected by monitoring the sensor's conductance.
Many efforts have been done to improve and develop the detection of single-electron charges.\cite{1998SchoelkopfScience, 2004VandersypenAPL, 2006QinAPL, 2009BarthelPRL, 2015OtsukaSciRep, 2020NoiriNanolett} For the high-speed measurement, an rf reflectometry, which monitors the rf signal reflected from a resonator containing the charge sensor, is a powerful tool to realize the fast readout of the charge states in quantum dots.\cite{1998SchoelkopfScience, 2006QinAPL, 2009BarthelPRL, 2015OtsukaSciRep, 2015StehliklPRAppl, 2020NoiriNanolett} Recently, it was reported that this method is applicable to not only gallium arsenide (GaAs) but also silicon (Si) quantum dots\cite{2020NoiriNanolett} by optimizing the device structure. 
The readout speed of single-electron charges is limited by the readout noise in the reflectometry. Some integration time is required to improve the signal-to-noise ratio (SNR) and achieve a high fidelity readout.\cite{2015OtsukaSciRep, 2020NoiriNanolett} 
Recently, to reduce the noise from the amplifier, an ultra low noise amplifier realized by a Josephson parametric amplifier (JPA) has been reported.\cite{2008YamamotoAPL} JPA could reduce the amplifier noise and high-speed readout of charge states is demonstrated.\cite{2015StehliklPRAppl} To proceed the improvement by reducing the noise in the whole system, investigating the noise in the system is still important. In the previous studies, the noise components are attributed as device flicker noise and circuit noise including amplifiers\cite{1998SchoelkopfScience,2018YonedaNnano}. In this study, we investigate detail of gate voltage dependence of the noise and compared to the charge sensor sensitivity including a sensor operation point on a Coulomb peak.
\begin{figure}
\includegraphics[width=9cm]{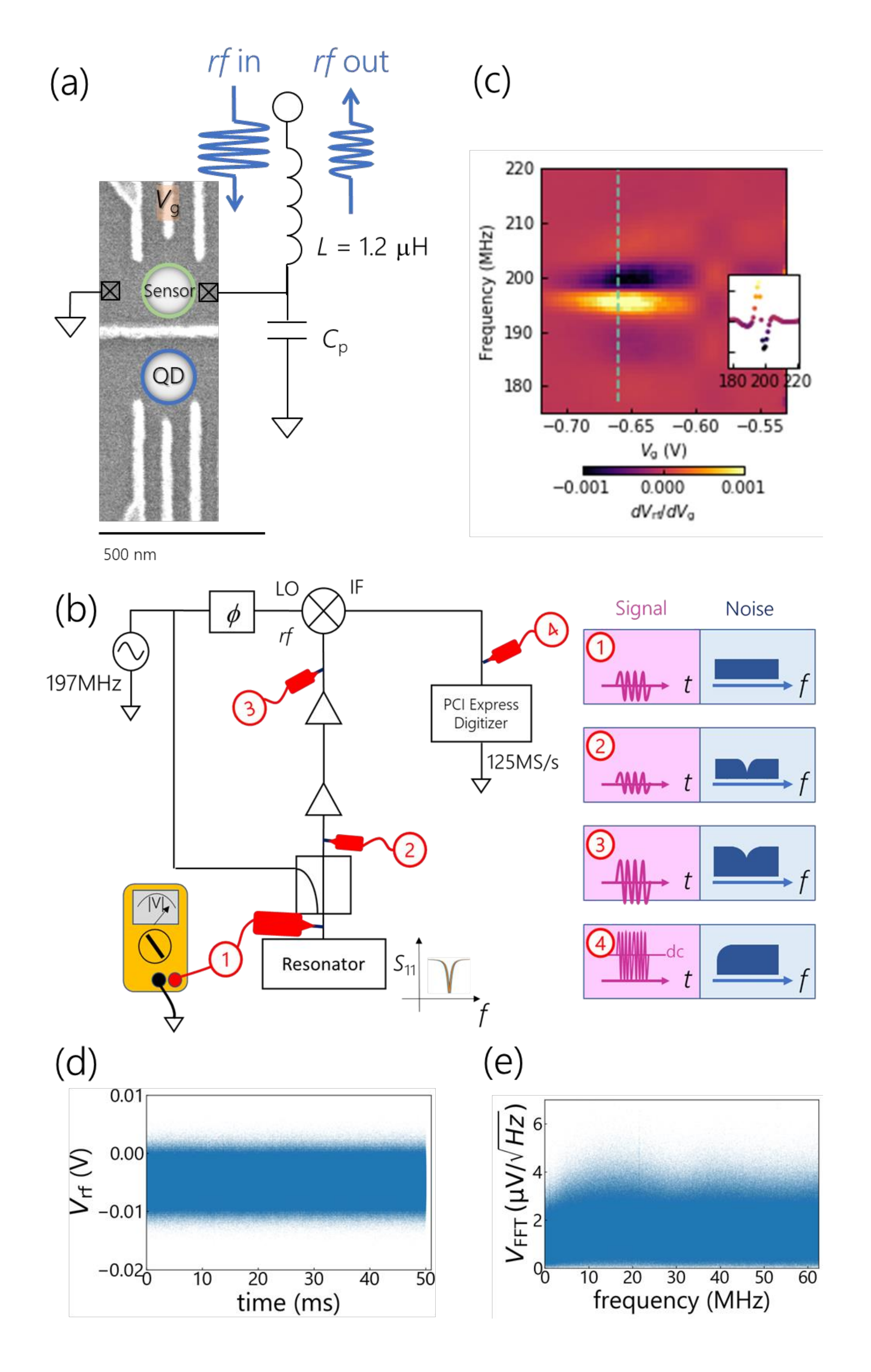}
\caption{(a) Schematic of the device structure and (b) measurement circuit. (c) Gate voltage dependence of the reflected rf signal. Inset indicates the signal along the dotted line at gate voltage $V_{\rm g}=$ -0.66 V. Note that the demodulated voltage depends on a phase difference $\delta \phi$ between the reflected signal and the local oscillator at the mixier as proportional to $\cos \delta \phi$. (d) The typical time resolved demodulated voltage and (e) its fast Fourier transform spectrum at $V_{\rm g} =$ -0.8785 V.}
\label{f1}
\end{figure}

Figure \ref{f1}(a) shows a scanning electron micrograph of the device and a schematic of the resonator. The device is fabricated on a GaAs/AlGaAs heterostructure wafer by depositing Ti/Au gate electrodes on the surface. The upper dot corresponds to the charge sensor and the bottom to the target quantum dots, respectively. The white structures correspond to the gate electrodes. The sensor quantum dot acting as variable resistance is embedded in an rf resonator circuit constructed by a chip inductor of 1.2 $\rm \mu \rm H$ and a stray capacitance $C_{\rm p}$ of 0.54 pF. The value of the inductance is determined by considering impedance matching treating the conductance of the charge sensor. The reflected rf signal from the resonator is modified depending on the sensor conductance. Figure \ref{f1}(b) shows the measurement setup of the rf reflectometry. An rf signal is applied to the resonator including the charge sensor through a directional coupler. The reflected rf signal is amplified and demodulated at the mixer by using the local oscillator. The down-converted signal voltage $V_{\rm rf}$ is digitized by a broadband digitizer. All  measurements in this paper are conducted in a dilution fridge at a temperature of 40 mK.

We measure $V_{\rm rf}$ as a function of the gate voltage $V_{\rm g}$ and rf carrier frequency. 
To extract the $V_{\rm g}$ dependence, a numerical derivative of $V_{\rm rf}$ with respect to $V_{\rm g}$ is evaluated as shown in Figure \ref{f1}(c) and the inset indicates characteristics at $V_{\rm g}$ = -0.66V.
Note that the change in the frequency direction is induced by an additional phase difference with changing the frequency, which is not compensated in this measurement.
The resonance is around 197 MHz, and the charge sensor has a maximum sensitivity at the frequency.
There is clear $V_{\rm g}$ dependence, which suggests that $V_{\rm rf}$ is very sensitive to the conductance of the charge sensor.
After fixing the frequency of the carrier rf signal, we optimize the phase shifter between the signal generator and mixer to achieve the maximum $V_{\rm rf}$.
We next monitor $V_{\rm rf}$ for 50 ms with every 8 ns time resolution corresponding to a sampling rate of 125MHz. 
Figure \ref{f1}(d) shows a typical time trace of $V_{\rm rf}$ at $V_{\rm g}=$ -0.8785 V. There is noise superimposed on the signal. 
We process the data by fast Fourier transform (FFT) as shown in Figure \ref{f1}(e). 
Note that the folding noise above Nyquist frequency of 62.5 MHz can be neglected due to the low pass filter before digitizing.
The FFT spectrum decreases below around 10 MHz. 
To explain this attenuation, we consider frequency responses of the noise passing through the resonator, the amplifier, and the mixier by assuming an input white noise as shown in figure \ref{f1}(b).
The possible candidates of the noise are thermal noise and emitted noise from the input port of the amplifier.
At the output port of the directional coupler, the noise is attenuated by the resonator around the resonance frequency. Next, the noise and the signal are increased by the amplifiers. In addition to amplification, amplifiers induces additional noise known as noise figure. 
The effect of the noise figure results in an offset value in the FFT spectrum of the readout noise. Finally, the mixer acts as a down-converter of the noise (and the signal) which demodulates 197 MHz to DC. Due to the bandwidth of the digitizer, we can neglect a second order harmonic component of 394 MHz.
Therefore, a frequency dependence of the noise before the mixer looks to be shifted from the resonance frequency to DC. In this wise, the observed frequency dependence of the noise shown in figure \ref{f1}(e) can be explained by considering the measurement circuit.
We summarize that the reflection property of the resonator around the resonance frequency will appear in the readout noise around DC by the down-conversion. The width of the decreased FFT spectrum shows good agreement with the width of the resonance. We also point out that the intrinsic noise of the amplifiers will result in the offset value in the FFT spectrum of the readout noise. 
\begin{figure}
\includegraphics[width=9cm]{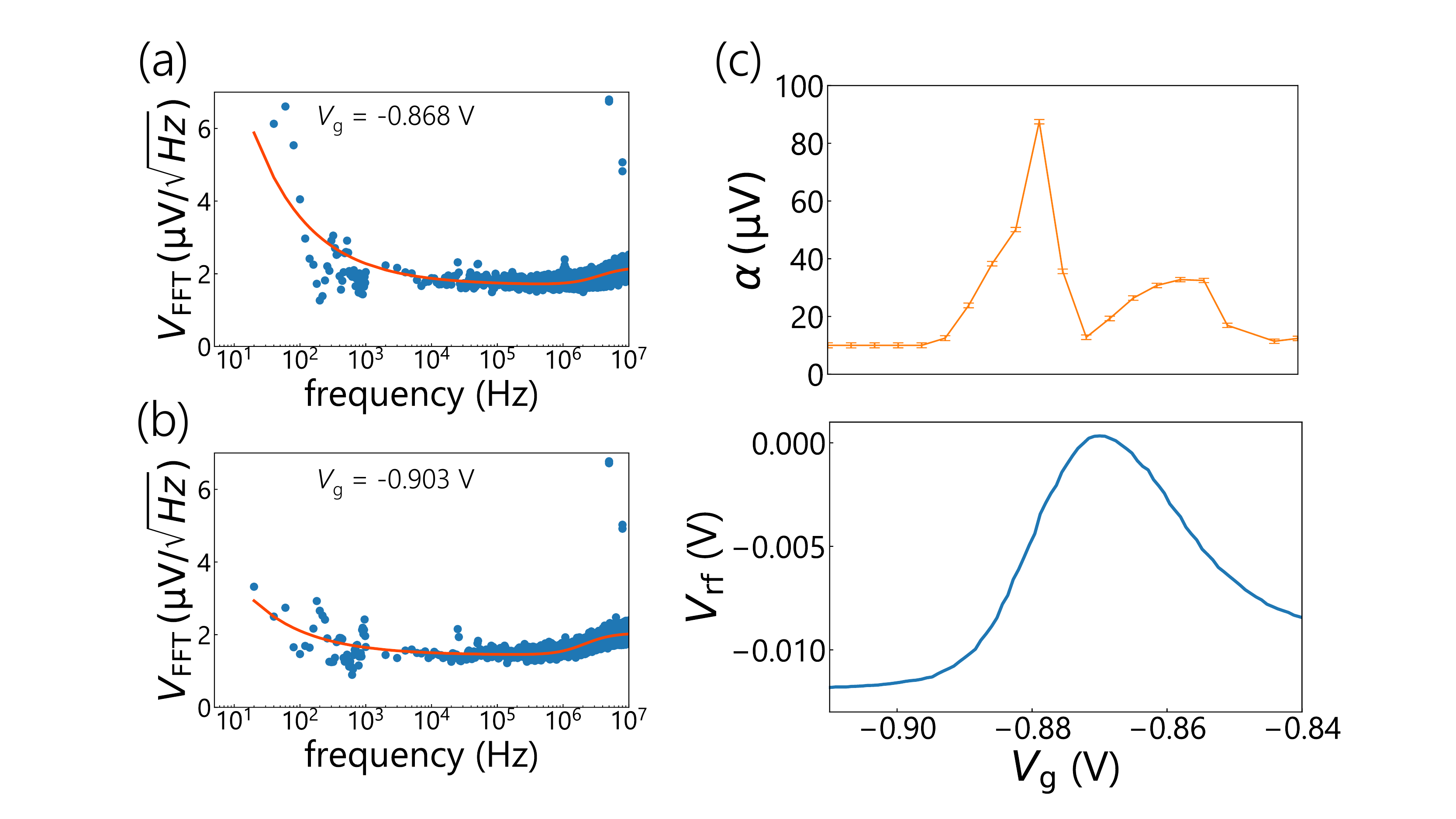}
\caption{Noise spectra at (a) gate voltage $V_{\rm g}$=-0.868V and (b) -0.903V. (c) $V_{\rm g}$ dependence of the flicker noise amplitude and the rf voltage.}
\label{f2}
\end{figure}

For further investigation of the noise, we plot the FFT spectrum with a logarithmic frequency axis in Fig. 2(a). 
We adapt the Bartlett method to reduce the fluctuation above 1 kHz.
We can observe a steep increase of the noise below 10 kHz. This can be understood as the carrier amplitude is modulated by the low-frequency device charge noise. The modulated $V_{\rm rf}$ is described by 
\begin{align}
    V_{\rm rf}= V_{\rm dc} \sum_{\omega} N(\omega)\sin \omega t
\end{align}
where $V_{\rm dc}$ is the dc component of $V_{\rm rf}$. $N(\omega)$ describes the frequency dependence of the device noise spectrum. Such flicker noise is reported in the previous studies.\cite{1998SchoelkopfScience, 2018YonedaNnano}
A possible candidate of the device noise is charge fluctuation noise of the trapping sites near the interface of the gate electrodes\cite{2004JungAPL, 2014PaladinoRMP} and DX centers\cite{1979LangPRB}. Charges trapped in the sites fluctuate randomly with time.\cite{1989KirtonAdvP}
We analyze total noise spectrum by the sum of the flicker noise and a symmetric Lorentz function $L_{\rm S}(f)$ describing the noise reduction below 10~MHz.
\begin{align}
    V_{\rm FFT}(f) = \frac{\alpha}{f^{\frac{1}{2}}} - L_{\rm S}(f) + \rm offset
\end{align}
Here, $\alpha$ is the amplitude of the flicker noise. The first term corresponds to the device noise, the second to the circuit noise from the performances of resonator and the amplifier, and the third to the intrinsic noise of the amplifier. Note that the $1/f$ function of the flicker noise is applicable to the power spectral density and that $V_{\rm FFT}(f)$ of the flicker noise has a $1/f^{\frac{1}{2}}$ dependence. The lines in Fig. 2(a) and (b) show the result of the fitting using Eq. (2). Those show good agreements with the data.
\begin{figure}
\includegraphics[width=7cm]{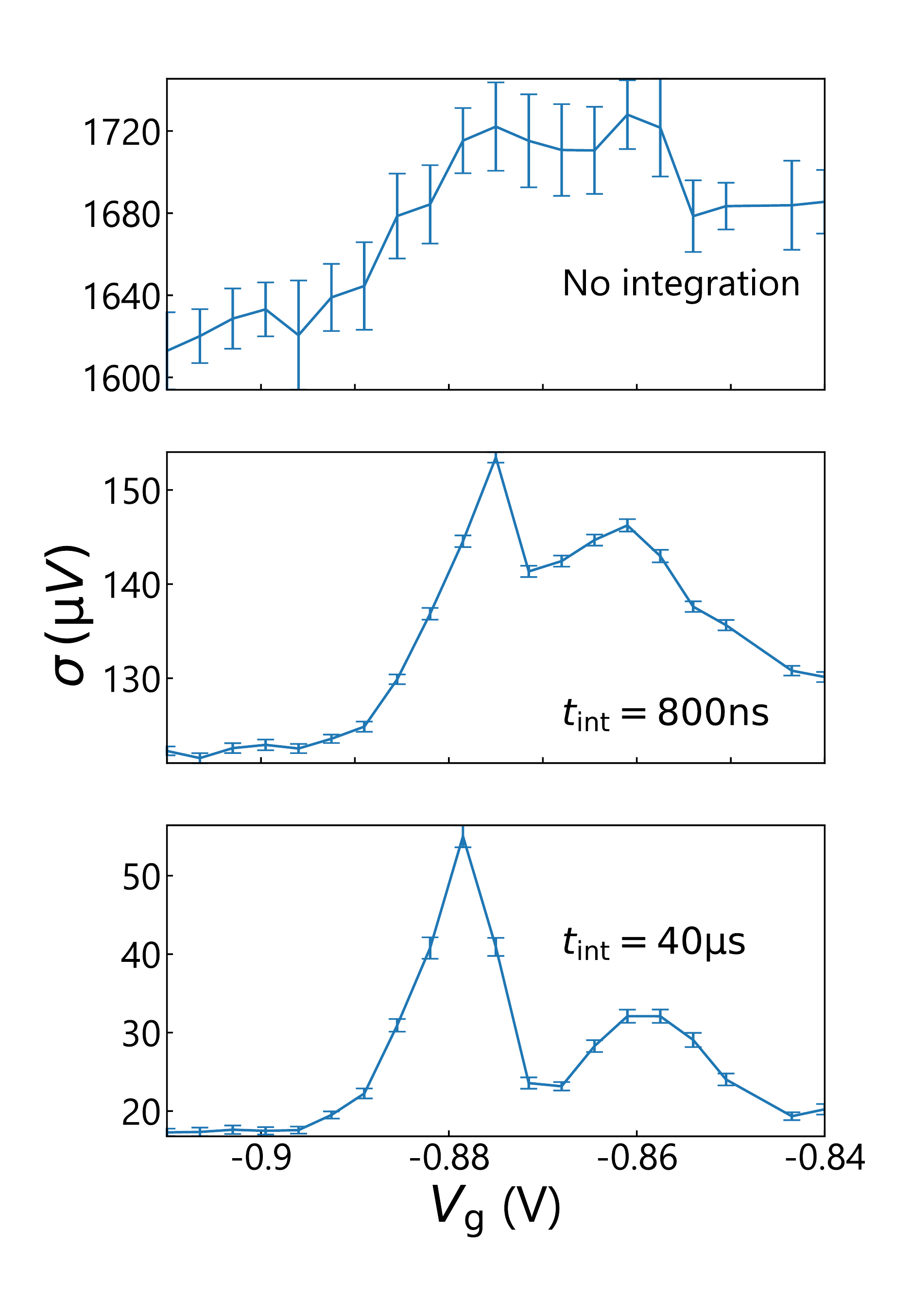}
\caption{The gate voltage dependence of the standard deviations of readout noise without and with integration times of 800 ns, and 40 $\rm \mu \rm s$.}
\label{f3}
\end{figure}
Furthermore, we evaluate $V_{\rm g}$ dependence of $\alpha$ shown in Fig. \ref{f2}(c). The y-axis of the top graph corresponds to $\alpha$ and the bottom one to $V_{\rm rf}$. The $V_{\rm g}$ dependence of $V_{\rm rf}$ corresponds to a single Coulomb peak of the sensor dot. The conductance of the sensor is around $0.6\,\frac{e^2}{h}$ at the Coulomb peak. On the other hand, there are two peaks in the dependence of $\alpha$. By comparing $V_{\rm g}$ dependence of $\alpha$ and $V_{\rm rf}$, we find out that $\alpha$ becomes large when the slope of $V_{\rm rf}$ becomes steep ($\left|\frac{dV_{\rm rf}}{dV_{\rm g}}\right|$ becomes large).
Indeed, $V_{\rm g}$ dependence of $\alpha$ nicely reproduces that of $\left|\frac{dV_{\rm rf}}{dV_{\rm g}}\right|$.
This is means $V_{\rm rf}$ becomes sensitive to the fluctuation of effective $V_{\rm g}$ induced by the device noise when $\left|\frac{dV_{\rm rf}}{dV_{\rm g}}\right|$ becomes large.
Althrough, this flicker noise is already reported in previous studies, we investigate the $V_{\rm g}$ dependence of the noise and find out the relationship between the noise and the sensitivity of the charge sensor in detail. Especially, we confirmed that the flicker noise takes a minimal value at the sensor Coulomb peak even with a finite sensor current.
As the conventional flicker noise from circuit components will not depend on $\left|\frac{dV_{\rm rf}}{dV_{\rm g}}\right|$, the flicker noise observed in this measurement is dominated by the device noise, suggesting that rf reflectmetry measurement is a very powerful tool to probe the pure device noise such as charge trapping sites near the interface in the heterostructure and DX centers.

Finally, we study the relationship between the readout noise and the measurement integration time. The standard deviations $\sigma$ by Gaussian fitting to the histograms of the processed data with the integration are evaluated. Figure \ref{f3} shows the $V_{\rm g}$ dependence of $\sigma$ with no integration (8 ns sampling) and with integration times $t_{\rm int}$ of 800 ns and 40 $\rm \mu \rm s$. The $V_{\rm g}$ dependence clearly changes from a single peak to double peaks with increasing integration time. In the case of short integration time, the single peak behavior originates from the the gate dependence of the conductivity, as well as the reflective property of the sensor dot because the effect of the noise reflected from the resonator is increased with the increase of the reflected signal. 
On the other hand, the double peak behavior in $t_{\rm int} = 40 \mu \rm s$ is similar to that in Fig. \ref{f2}(c). With the increase of the integration time, the device noise becomes dominant in $\sigma$. The time resolution of 40 $\mu \rm s$ corresponds to a Nyquist frequency of 12.5 kHz, which is in good agreement with the device noise dominated region. Note that the device noise in the Coulomb blockade regime should be zero because $V_{\rm rf}$ has no sensitivity to $V_{\rm g}$. However, the values of $\sigma$ in the blocked region $V_{\rm g}$ = -0.91 to -0.90 remain finite values. This offset value corresponds to the circuit noise. Therefore, the effect of the amplifier always contributes to $\sigma$, and it is effective to improve the noise figure of the amplifier for high-speed and accurate state estimation as already demonstrated in fast measurements of cavity-coupled quantum dots.\cite{2015StehliklPRAppl}
However, the origin of $\sigma$ changes from the circuit noise to the flicker noise by time integration and the effect of the flicker noise is much rather than the amplifier noise with long integration time.

In summary, we investigate the gate voltage dependence of electrical readout noise in high-speed rf reflectometry using a GaAs quantum dot. We measured detail of the flicker noise and find out the relationship between the noise and sensitivity of the charge sensor.
Especially, we confirmed that the flicker noise takes a minimal value at the sensor Coulomb peak.
We also point out that the device noise becomes dominant when we increase the measurement integration time to reduce the effect of the noise. 
The source of noise distribution is changed from the circuit noise to the flicker noise with the increase of time integration.
These findings are important in improving the measurement setup of the rf reflectometry. Also, rf reflectometry can serve as a sensitive probe of the device noise, which induces decoherence of the qubits.

\section{Acknowledgements}

The authors thank for T. Kumasaka and T. Aizawa for their technical supports and fruitful discussions.
Part of this work is supported by
PRESTO (JPMJPR16N3),
JST, MEXT Leading Initiative for Excellent Young Researchers,
Grants-in-Aid for Scientific Research from JSPS (Nos. 20H00237, 20J14418),
Takahashi Industrial and Economic Research Foundation Research Grant,
The Murata Science Foundation Research Grant,
Samco Foundation Research Grant,
Casio Science Promotion Foundation Research Grant,
The Thermal \& Electric Energy Technology Foundation Research Grant,
Telecom Advanced Technology Research Support Center Research Grant,
Izumi Science and Technology Foundation Research Grant,
JFE 21st Century Foundation Research Grant,
and FRiD Tohoku University.
M.S. acknowledge financial supports from JST-OPERA and DIARE of Tohoku University.
A.D.W. and A.L. acknowledge gratefully support of DFG-TRR160, BMBF-Q.Link.X  16KIS0867, and the DFH/UFA CDFA-05-06.

\end{document}